\documentclass[aps,nofootinbib,superscriptaddress,twocolumn,preprintnumbers]{revtex4}
\usepackage{graphicx,amsmath,amssymb,amsbsy,latexsym,verbatim}
\usepackage{mathpazo}

\usepackage{setspace}
\usepackage{amsmath,amssymb,amsfonts,amsthm}
\usepackage{graphicx}
\usepackage{enumerate}

\newcommand{\f}{\frac}
\newcommand{\tf}{\tfrac}

\newcommand{\dd}{\mathrm{d}}

\def\tm{t}

\def\ep{\varepsilon}

\newcommand{\be}{\nopagebreak[3]\begin{equation}}
\newcommand{\ee}{\end{equation}}
\newcommand{\ba}{\nopagebreak[3]\begin{eqnarray}}
\newcommand{\ea}{\end{eqnarray}}
\newcommand{\nn}{\nonumber \\}
\newcommand{\bc}{}

\newcommand{\bra}[1]{\ensuremath{\,\langle\,{\tiny #1}\,|\,}}
\newcommand{\ket}[1]{\ensuremath{\,|\,{#1}\,\rangle\,}}

\newcommand{\bk}[3]{{\langle\,{ #1}\,|#2|\,{#3}\,\rangle\,}\!}
\newcommand{\sm}[1]{ \left( \begin{smallmatrix} #1 \end{smallmatrix}\right)}

\begin{document}
\title{ \Large On the spinfoam expansion in cosmology}

     \author{Carlo Rovelli}      
     \affiliation{Centre de Physique Th\'eorique de Luminy\footnote{Unit\'e mixte de recherche (UMR 6207) du CNRS et des Universit\'es de Provence (Aix-Marseille I), de la M\'editerran\'ee (Aix-Marseille II) et du Sud (Toulon-Var); laboratoire affili\'e \`a la FRUMAM (FR 2291).} , Case 907, F-13288 Marseille, EU}     
     \author{Francesca Vidotto${}^{1,}$}
     \affiliation{Dipartimento di Fisica Nucleare e Teorica, Universit\`a degli Studi di Pavia \\ and Istituto Nazionale        di Fisica Nucleare, Sezione di Pavia, via A. Bassi 6,        27100 Pavia, EU}

\date{\small \today}
\begin{abstract}
\noindent
We consider the technique introduced in a recent work by Ashtekar, Campiglia and Henderson, which generate a spinfoam-like sum from a Hamiltonian theory. We study the possibility of using it for finding the generalized projector of a constraint on physical states, without first deparametrising the system. We illustrate this technique in the context of a very simple example. We discuss the infinities that appear in the calculation, and argue that they can be appropriately controlled. We apply these ideas to write a spinfoam expansion for the ``dipole cosmology".

\end{abstract}

\maketitle

\section{Introduction}

The spinfoam formalism \cite{sf} and canonical loop quantum gravity \cite{lqg,libro} can ideally be viewed as the covariant and canonical versions  of a background-independent quantum theory of gravity, a scenario nicely realized in three dimensions \cite{Noui:2004iy}. In a recent paper \cite{Ashtekar:2009dn}, Ashtekar, Campiglia and Henderson find an ingenious way to implement this scenario in the context of Loop Quantum Cosmology, by introducing a technique for translating the dynamics of the quantum theory into a spinfoam-like sum-over-histories dynamics.   

Here we apply this technique to the direct calculation of the projector on the kernel of a constraint, and we illustrate it the context of an extremely simple example.   The exercise allows us to discuss the appearance of certain infinities. We argue that these are spurious and can be properly dealt with. 

We use these results to write a spinfoam formulation for the ``dipole" cosmology recently introduced in  \cite{Rovelli&Vidotto1}. This can be done in a fully relational way; without the need of deparametrising the system using a scalar field.

\subsection{The Ashtekar-Campiglia-Henderson expansion}

The idea of the Ashtekar, Campiglia and Henderson (ACH) expansion \cite{Ashtekar:2009dn} can be resumed as follows.  Consider a self-adjoint operator $C$ on a Hilbert space $\cal H$, where a discrete orthonormal basis $\ket x$ is given.  We are interested in the evolution operator $e^{i\tm C}$.  As in the standard Feynman construction of the path integral,we write
\be
e^{i\tm C}= \underbrace{e^{i\ep C}...e^{i\ep C}}_{N\ {\rm times}}
\ee
where $N$ is an arbitrary positive number and $\ep\equiv t/N$. In this way, the evolution in the interval $[0,t]$ is realized by chopping the interval in $N$ steps and evolving one step at the time.  Inserting $N-1$ completeness this reads 
\ba
e^{i\tm C}&=&
\sum_{x_1 ... x_{N-1}}e^{i\ep C} \ket{x_{N}}
 \bk{x_{N-1}}{e^{i\ep 
C}}{x_{N-2}}
 \\ \nonumber
 && \hspace{7em} ... \bk{x_2}{e^{i\ep 
C}}{x_{1}} \bra{x_{1}} e^{i\ep 
C}
\ea
Therefore the matrix elements of the evolution operator can be written in the form 
\be\label{amplitude}
\bk{x'}{e^{i\tm C}}{{x}}=
\sum_{x_1 ... x_{N-1}}
P_{{x'},x_{N-1}}...P_{x_1,{x}}   \ .
\ee
where 
\be
P_{x,x'}=\bk{x}
{e^{i \ep C}}{x'}    \ .
\ee
That is, the matrix elements of the evolution operator can be expressed in terms of a sum over ``histories" $(x_{N-1}, ... , x_1)$, with an amplitude $A(x_{N-1}, ... , x_1)=
P_{{x'},x_{N-1}}...P_{x_1,{x}}$ associated at each history. 
Since the expression \eqref{amplitude} is true for any $N$, we can equally write
\be
\bk{x'}{e^{itC}}{{x}}=\lim_{N\to\infty}\sum_{x_1 ... x_{N-1}}P_{{x'},x_{N-1}}...P_{x_1,{x}}  \ .
\ee
Let us write the matrix elements of $C$ as $C_{xx'}\equiv\bk x C{x'}$, and the diagonal ones as  $C_x\equiv C_{xx}\equiv\bk x C{x}$.
Then, for $N$ large enough (that is, $\ep$ small), we can write that 
\be
\bk{x}{e^{i\ep C}}{x}\sim e^{i\ep C_{x}}
\ee
while if $x\ne x'$
\be
\bk{x}{e^{i\ep C}}{x'}\sim i\ep C_{xx'}   \ ,
\ee
up to terms of higher order in $1/N$. The key idea of Ashtekar-Campiglia-Henderson is now to rearrange the sum over histories by collecting together all histories where there is a number $M$ of ``jumps". A jump being a step of the history where $x_{n+1}\ne x_n$. 
For each history, let $N_m$ be the position of the $m$-th jump. Then we have immediately 
\ba
\bk{x'}{e^{itC}}{{x}}&\!\!\!=&\!\!\!\lim_{N\to\infty}\sum_{M=0}^N 
\sum_{N_{M}...N_1}
\sum_{x_{M-1} ... x_1}
\!\!\!\!\!  i\ep C_{{x'}x_{M-1}}...
 i\ep C_{x_1 {x}}\nonumber\\ 
&& (e^{i\ep C_{x'}})^{N-N_{M-1}}
...
(e^{i\ep C_{{x}}})^{N_1}
\ea
where the $x_m$'s are the $M-1$ intermediate values that the history takes, each lasting for $N_{m+1}-N_{m}$ steps.  The only subtlety above is to keep truck of the limits of the sums over the $N_m$'s, which are not indicated above.  A moment of reflection shows that these are 
\be
\sum_{N_{M}...N_1} = 
\sum_{N_{M}=M}^{N-1}\, \sum_{N_{M-1}=M-1}^{N_M-1}\, \ldots\, \sum_{N_1=1}^{N_2-1}\,\,
\ee
the range of the sum being determined by the fact that $N<N_M<N_{M-1}<...<N_2<N_1$ where all the $N_m$ are different from zero. Finally, the last step of the Ashtekar-Campiglia-Henderson construction is to take the continuum limit of these sums. This can be done by noticing that the $N\to\infty$ limit of the sums is nothing else than the definition of the Riemann integral. Thus for instance the first sum in $N_M$ converges to an integral from zero to $t$, and so on.  This gives
\ba
\sum_{N_{M}=M}^{N-1}\ep \, && 
\sum_{N_{M-1}=M-1}^{N_M-1}\ep \, \ldots\, \sum_{N_1=1}^{N_2-1}\ep \,\,
\nonumber \\  \label{sums}
&&\to \
\int_0^t  \dd\tm_M \ \int_0^{t_M} \dd\tm_{M-1} \ ... \ \int_0^{t_2} \dd\tm_1  \ . 
 \ea
 where $t_m=\varepsilon N_m$. 
Bringing altogether, we have
\ba\label{eccola}
\bk{x'}{e^{itC}}{{x}}&=&\lim_{N\to\infty}\sum_{M=0}^N 
\sum_{x_1 ... x_{M-1}}
\left(C_{{x'}x_{M-1}}...C_{x_1 {x}}\right)\nn
&&  \mu(x_1, ...,x_{M-1},t)
\ea
where 
\ba\label{eccola2}
\mu(x_1, ..., x_{M-1},t)&&=  i^{\scriptscriptstyle M}
\int_0^t \dd\tm_M  ...  \int_0^{t_2} \dd\tm_1 \nn
&&(e^{i C_{x'}})^{t-t_{M}}
...
(e^{i C_{{x}}})^{t_1}\ .
\ea
The integrals can be performed explicitly. If all $C_x$ are different, for instance, this gives
\ba\label{mu}
\mu(x_1, ..., x_{M-1},t)=\sum_{m=0}^M\frac{e^{itC_{x_mx_m}}}{\prod_{m'\ne m}(C_{x_mx_{m}}-C_{x_{m'}x_{m'}})}\ .
\ea
The expression can easily be generalized to cases where some $C_x$'s are equal (an example is below).

\subsection{The Projector}

Suppose now that we are interested in solving the equation 
\be
\label{Cpsi}
C\ket\psi=0   \ .
\ee
The solutions of this equation are in $\cal H$ if zero is a discrete eigenvalue of $C$ (case $(i)$); while they are in a suitable extension $\cal K$ of  $\cal H$ if zero belongs to the continuous spectrum of $C$ (case $(ii)$).  Following a procedure now common in quantum gravity, instead of searching directly for the $\ket\psi$ states (or generalized states in $\cal K$) that solve (\ref{Cpsi}), we search for the operator $P$ that maps $\cal H$ to the space of the solutions.  In case $(i)$, the operator is a projector which can be formally written as
\be\label{proggy}
P=\lim_{T\rightarrow\infty}\tfrac1{2T}\int^{+T}_{-T}
\dd\tm ~ e^{i\tm
C}
\ee
or, alternatively
\be\label{proggy22}
P=\lim_{T\rightarrow\infty} e^{-T
C}.
\ee
It is easy to see that these operators project on the solutions of \eqref{Cpsi}, for instance by simply going to a basis that diagonalizes $C$. 
In case $(ii)$, $P$ can be written in the form
\be\label{proggy2}
P=\int^{\infty}_{-\infty}
\dd\tm ~ e^{i\tm
C}. 
\ee
Notice that if \eqref{Cpsi} is the Hamiltonian constraint of a background independent quantum theory, then  the integration variable $t$ in equations \eqref{proggy} and \eqref{proggy2} is not a physical time variable: rather, it is an unphysical coordinate time, or, equivalently, a ``group averaging" parameter, which will not appear in the physical transition amplitudes. We focus here on the discrete case $(i)$ for simplicity.  Inserting the ACH expansion \eqref{eccola} into (\ref{proggy}), we have
\ba\label{proggy333}
\bk{x'}{P}{{x}}&=&\lim_{T\rightarrow\infty}\tfrac1{2T}\int^{+T}_{-T}
\dd\tm ~ \lim_{N\to\infty}\sum_{M=0}^N 
\sum_{x_1 ... x_{M-1}}\\ \nonumber
&&\left(C_{{x'}x_{M-1}}...C_{x_1 {x}}\right)
  \mu(x_1, ...,x_{M-1},t).
\ea
If we exchange the two limits, we see immediately that a problem appears: the integral of \eqref{mu} in $\dd\tm$ behaves very badly.

\newpage
\section{A simple example}

Let us now compute explicitly the above expressions in a very simple case.  We take ${\cal H}=R^2$ and
\be\label{simham}
C = \sm{1&1\\ 1&1}
\ee
The solution of equation \eqref{Cpsi} is 
$
\ket{\psi}=\sm{~~1\\-1}
$
since one can easily see that $ \sm{1&1\\ 1&1}\sm{~~1\\-1}=\sm{0\\0}$. The orthogonal matrix $U=U^{-1} = \tf1{\sqrt{2}}   \sm{1&~~1\\ 1&-1}$ diagonalizes $C$:
\be
U\,C\,U^{-1}  =  \tf1{\sqrt{2}}    \sm{1&~~1\\ 1&-1} 
                                \sm{1&1\\ 1&1}
                                \tf1{\sqrt{2}}    \sm{1&~~1\\ 1&-1}
                        =   \sm{2&0\\ 0&0}
\ee
We call this matrix $D$. The projector on the solutions is
\be\label{prj}
P= \tfrac12    \sm{~~1&-1\\ -1&~~1}\ . 
\ee
%
%
%
Since zero is a discrete eigenvalue of $C$, we are in case $(i)$, and $P$ should be given by \eqref{proggy}.  To show that this is correct, let's compute the exponent. This is, by definition,
\ba
e^{i\tm
C}&=&U\,
e^{i\tm
D}
\,U^{-1} 
\\ &=&
 \tf1{\sqrt{2}}    \sm{1&~~1\\ 1&-1} 
\   \sm{e^{i2\tm}&0\\ 0&0}  \
  \tf1{\sqrt{2}}    \sm{1&~~1\\ 1&-1} 
\nn &=&
\tf12\sm{ e^{i2\tm}+1~ & ~ e^{i2\tm}-1 \\   e^{i2\tm}-1~ & ~  e^{i2\tm}+1}.
\nonumber
\ea
Inserting this in  \eqref{proggy} gives 
\ba
P&=&\lim_{T\rightarrow\infty}\tfrac1{2T}\int^{+T}_{-T}
\dd\tm 
\tf12\sm{ e^{i2\tm}+1~ & ~ e^{i2\tm}-1 \\   e^{i2\tm}-1~ & ~  e^{i2\tm}+1}
\nn &=&
\lim_{T\rightarrow\infty}\tfrac1{2T}\int^{+T}_{-T}
\dd\tm ~
[\tf12 e^{i2\tm}C + P]
\ea
The first term, once integrated, gives zero in the limit $T\rightarrow\infty$, so we are left just with the second term that gives:
\be
P=\
\lim_{T\rightarrow\infty}\tfrac1{2T}\int^{+T}_{-T}
\dd\tm ~
\tf12 \sm{~~1&-1\\ -1&~~1} 
=
\tf12 \sm{~~1&-1\\ -1&~~1}
\ee
which is indeed \eqref{prj}.  Let us now chose the basis $\ket x$, with $x=\pm1$, where $\ket{+1}=\sm{1\\0}$ and $\ket{-1}=\sm{0\\1}$ for writing the sum over histories. 

We consider sequences $\sigma$ of state of the kind $\ket{x_n}=\ket{\pm 1}$.
As in the previous section, we can write
\ba
P&=&\lim_{T\rightarrow\infty}\tfrac1{2T}\int^{+T}_{-T}
\dd\tm\left( e^{i\tf\tm{N}
C}\right)^N
\\ &=&
\lim_{T\rightarrow\infty}\tfrac1{2T}\int^{+T}_{-T}
\dd\tm
\sm{ \tf{e^{i2\varepsilon}+1}2~ & ~ \tf{e^{i2\varepsilon}-1}2 \\   \tf{e^{i2\varepsilon}-1}2~ & ~ \tf{e^{i2\varepsilon}+1}2}^N
\ea
where we have called $\f\tm{N}=\varepsilon$.
We consider now the transition amplitude
\ba
\bk{\pm 1}{P}{\pm 1}\!\!
&=& \!\!
\lim_{T\rightarrow\infty}\tfrac1{2T}\int^{+T}_{-T}
\!\!
\dd\tm 
\bk{\pm 1}{\!
\sm{ \tf{e^{i2\varepsilon}+1}2~ & ~ \tf{e^{i2\varepsilon}-1}2 \\   \tf{e^{i2\varepsilon}-1}2~ & ~ \tf{e^{i2\varepsilon}+1}2}^{\!\!\!N}\!\!\!
}{\pm1}
\nonumber
\ea
We insert $N-1$ completeness in the product of matrices. This gives rise
to a sum over the possible sequences  $\sigma$, alternating between the two states $\ket{\pm 1}$.  Each time the history steps from a plus to a minus, or from a minus to a plus, we get a contribution $\tf{e^{i2\varepsilon}-1}2$, while for each step where the sign does not change we get a contribution $\tf{e^{i2\varepsilon}+1}2$. Therefore
\ba\bk{\pm 1}{P}{\pm1}\!\!&=&
\sum_{  \sigma }
\lim_{T\rightarrow\infty}\tfrac1{2T}\int^{+T}_{-T}
\dd\tm 
(\tf{e^{i2\varepsilon}+1}2)^{N-M}
(\tf{e^{i2\varepsilon}-1}2)^{M}
\nonumber
\ea
where $M$ is the number of times the sequence changes sign. The exponents gives 
terms of the form $\sum_k c_k e^{i2k\ep}+1$, and we know that the exponents are suppressed by the integration and the limit. Thus we 
obtain
\ba\bk{\pm 1}{P}{\pm1}\!\!&=&
\sum_{  \sigma }
\lim_{T\rightarrow\infty}\tfrac1{2T}\int^{+T}_{-T}
\dd\tm 
\left(\f{1}{2}\right)^{N-M}\left(-\f{1}{2}\right)^{M}
\nn &=&
\sum_{  \sigma}
\lim_{T\rightarrow\infty}\tfrac1{2T}\int^{+T}_{-T}
\dd\tm 
\left(\f{1}{2}\right)^{N}\left(-1\right)^{M}\nn &=&
\f{1}{2^N}\sum_{ \sigma}
\left(-1\right)^{M} 
\ea
Now the number of changes must be even if the initial and final states are the same and odd otherwise. Therefore 
\ba\label{ecco}
\bk{\pm 1}{P}{\pm1}\!\!=\pm\pm\f{1}{2^N}\sum_{  \sigma} 1 
\ea
where ${\cal N}=\sum_{  \sigma} 1$ is the number of sequences of length $N$, with fixed beginning and starting point.  Easily, ${\cal N}=2^{N-1}$, so that we obtain 
\ba
\bk{\pm 1}{P}{\pm1}\!\!=\pm\pm\f{1}{2}
\ea
which is, once more, precisely the projector \eqref{prj}. Therefore the matrix elements of the projector \emph{can} be written as sums over histories. 

So far, we have \emph{not} applied the ACH expansion in the number of jumps.  Let us do so now.
From \eqref{eccola} and \eqref{eccola2}, we have
\be\label{eccola11}
\bk{x'}{e^{itC}}{{x}}=\lim_{N\to\infty}\sum_{M=0}^N 
\sum_{x_1 ... x_{M-1}}
 \mu(x_1, ...,x_{M-1},t)
\ee
where 
\ba\label{eccola12}
\mu(x_1, ..., x_{M-1},t)= i^{\scriptscriptstyle M}
\int_0^t \dd\tm_M  ...  \int_0^{t_2} \dd\tm_1 \  
(e^{i\f t N})^{N-M}.
\ea
which in the large $N$ limit will give
\ba\label{eccola12bis}
\mu(x_1, ..., x_{M-1},t)=e^{it} \f{(it)^M}{M!}.
\ea
Fix the initial and final states, for instance say they are the same. Then notice that there is a single possible sequence with $M$ steps alternating pluses and minuses if $M$ is even, and none if $M$ is odd. Vice-versa,  if the initial and final values are different, then there is a single possible sequence if $M$ is odd, and none if $M$ is even.
Therefore the sum over histories gives unit for even $M$ and zero for odd $M$ for equal boundary states, and vice-versa for different boundary states. Bringing all together, we have
\ba\label{eccola111}
\bk{+}{e^{itC}}{+}&=&\lim_{N\to\infty}e^{it}\sum_{even\ M=0}^N 
 \f{(it)^M}{M!}.
\\
\bk{+}{e^{itC}}{-}&=&\lim_{N\to\infty}e^{it}\sum_{odd\ M=0}^N 
 \f{(it)^M}{M!}.
\ea
Inserting this in the definition of the projector gives
\ba\label{key}
\bk{+}{P}{+}&=&\lim_{T\to\infty}\f{1}{2T}\int_{-T}^T \dd t
\lim_{N\to\infty}e^{it}\!\!\sum_{even\ M=0}^N 
 \f{(it)^M}{M!},\ \ \ \ \ \ \  \\  \label{keykey}
\bk{+}{P}{-}&=&\lim_{T\to\infty}\f{1}{2T}\int_{-T}^T \dd t
\lim_{N\to\infty}e^{it}\!\!\sum_{odd\ M=0}^N 
 \f{(it)^M}{M!}.
\ea
This is our key result. Let us discuss it.  If we take the $N$ limit first, we have easily
\ba\label{eccola111bis}
\bk{+}{P}{+}&=&\lim_{T\to\infty}\f{1}{2T}\int_{-T}^T \dd t\ 
e^{it}\cos(t)=\f{1}{2};\\
\bk{+}{P}{-}&=&\lim_{T\to\infty}\f{1}{2T}\int_{-T}^T \dd t\ 
e^{it}\sin(t) = -\f{1}{2},
\ea
which is, once again, the right result. This shows that the limits taken above, such as the expansion of the matrix elements in $\epsilon$, the replacement of the sum with integrals and the elimination of $M$ in the exponent in \eqref{eccola12} are legitimate.
But if we now take the $T$ limit before the $N$ limit in (\ref{key}--\ref{keykey}), we have a sum of divergent terms.  Thus, the ACH procedure is viable in this case for computing the matrix elements of the projector, but only provided that the $N$ limit is performed before the $T$ limit.

This may appear to be a serious problem at first sight, because the objective of the ACH expansion is to have a sum over $M$ that can be understood as a perturbative expansion. But the individual terms of the sum over $M$ diverge in the $T$ limit. 
\vskip3em
\section{The regularized projector}

There is a simple way to circumvent the problem described at the end of last section.  In most physical applications of a highly nontrivial theory, we are not interested in the exact result, but in results obtained in some approximation scheme, such for instance a perturbation theory.
Then we are only interested in solving the equations of the theory with some given margin of error. Thus, we may be content of computing an operator which ``almost" projects on the solutions of \eqref{Cpsi}. Say we fix a small number $\delta$ and search for the regularized projector 
\be
\label{proggydelta}
P_\delta=\tfrac{\delta}{2}\int^{\f{1}\delta}_{-\f{1}\delta}
\dd\tm ~ e^{i\tm C}
\ee
It is easy to see that this gives an approximate solution of \eqref{Cpsi}, in the sense that it projects on states $\ket\psi$ such that the norm of $C\ket\psi$ is smaller than $\delta$ times the norm of $\ket\psi$. In the basis that diagonalizes $C$, we have in fact
\be
P_\delta=\tfrac{\delta}{2}\int^{\f{1}\delta}_{-\f{1}\delta}
\dd\tm ~ e^{i\tm \sm{2 & 0\\0&0}}= \sm{\f\delta2\sin\f2\delta & 0\\0&1}=P+O(\delta).
\ee
Notice that the $\delta\to 0$ limit of $P_\delta$ is well defined, but $P_\delta$ is not analytic in $\delta=0$.  Using the above results, we can write 
\be\label{key2}
\bk{+}{P_\delta}{+}=\f{\delta}{2}\int^{\f{1}\delta}_{-\f{1}\delta}
 \dd t
\lim_{N\to\infty}e^{it}\!\!\sum_{even\ M=0}^N 
 \f{(it)^M}{M!}.
\ee
The integral can now be easily performed, giving
\be\label{key3}
\bk{+}{P_\delta}{+}=\lim_{M_{\rm max}\to\infty}\sum_{even\ M=0}^{M_{\rm max}} \f{ i^{\scriptscriptstyle M}}{M!} \ c_M.
\ee
where
\be
 c_M=\f{\delta}{2}\int^{\f{1}\delta}_{-\f{1}\delta}
 \dd t \  e^{it}\ t^M.
 \ee
For any finite $\delta$, the series in $M$ is convergent. In fact, since the $c_M$'s grow in $M$ as $(1/\delta)^{M+1}$, the sum is strongly suppressed by the factorial beyond a certain $M_{max}$. See the examples of Figure 1. 
\begin{figure}[h]
\includegraphics[height=2.5cm]{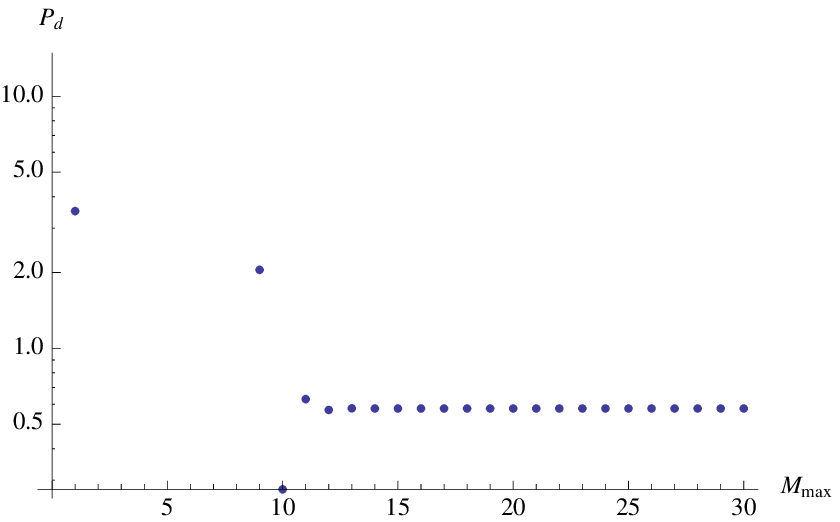}
\includegraphics[height=2.5cm]{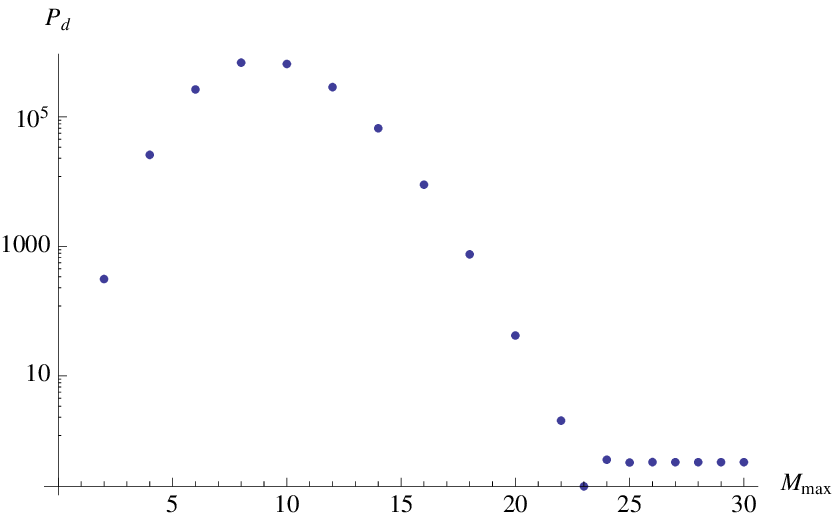}
\caption{The convergence of the series $\eqref{key3}$ for a matrix element of $P_\delta$ (whose exact value in the $\delta\to 0$ limit is $0.5$). For $\delta=0.1$ (left) the sum converges to $0.58$ in about 13 steps. For $\delta=0.05$, (right) the sum converges to $0.46$ in about 25 steps. }
\end{figure}

Therefore we can compute the projector $P$ to any given desired accuracy $\delta$ by means of a sum over histories where long histories do not matter. 

Let us now go back to the general problem. The example shows that the ACH expansion can be viable to compute $P$, but spurious infinities appear if the $T$ limit is taken before the $N$ limit. To solve this difficulty, we can compute a regularized version of $P$. This is given by 
\ba\label{proggyc}
\bk{x'}{P_\delta}{{x}}&=&\sum_{M=0}^\infty
\sum_{x_1 ... x_{M-1}} C_{{x'}x_{M-1}}...C_{x_1 {x}}
\\ \nonumber
&& \ \ \ \ 
  \mu_\delta(x_1, ...,x_{M-1}),
\ea
where 
\ba\label{eccola222}
\mu_\delta(x_1, ..., x_{M-1})&&= i^{\scriptscriptstyle M}\ 
\tfrac{\delta}{2}\int^{1/\delta}_{-1/\delta}
\dd\tm \
\int_0^t \dd\tm_M  ...  \int_0^{t_2} \dd\tm_1 \nn
&&(e^{i C_{x'}})^{t-t_{M}}
...
(e^{i C_{{x}}})^{t_1}\ .
\ea
For any given value of the regulator $\delta$, the sum over the length of the histories that defines the ACH expansion converges. This opens the possibility of computing $P$
to any given desired accuracy $\delta$ by means of a sum over histories where long histories do not matter. 

\vskip3em
\section{Dipole cosmology}

 The key result of  \cite{Ashtekar:2009dn} is the construction of a Feynman-like sum-over-paths formulation of the quantum dynamics of a FRW cosmology, where the single histories are sequences of an (arbitrary, but) finite number of discrete steps and there is no time-parameter to keep track of the duration of each single step.  This ``topological" character of the sum over paths can be seen as a realization of the background independence of the theory, and represents the remnant in cosmology of the full background independence of the spinfoam theory.  
 
 This  ``topological" character of the final sum over paths, on the other hand, is realized in a rather roundabout manner in  \cite{Ashtekar:2009dn}.  The FRW model is coupled to a scalar field.  Then, this field is taken as the independent time variable; that is, the evolution is resolved with respect to it. Finally, the intermediate scalar-field-time dependence is integrated away, leaving just a sum over histories of the purely gravitational variable, at given boundary values of both this variable and the scalar field. 

Can the same technique be applied in a more straightforward manner, without the need of singling out a time variable and deparametrize the system?  In principle, this can be done by applying the ACH expansion to the (generalized) projector $P$ on the solutions of the Wheeler-deWitt equation of a covariant theory. (A first attempt in this direction is in Section V of  \cite{Ashtekar:2009dn}.)  This is precisely what we have done in the previous section. Here, we apply the results of the previous sections, in a context where we do not introduce an effective scalar field and we do not deparametrize the evolution.  For this, we need to work with a cosmological model with a sufficient number of degrees of freedom for the relative evolution among variables to be defined. 

We consider the ``dipole cosmology" introduced in \cite{Rovelli&Vidotto1}.   This system represents the lowest order of a ``triangulated quantum cosmology". As shown in \cite{antonino}, it describes a Bianchi IX closed universe, plus a small amount of inhomogeneity.  The Hilbert space of the model admits a discrete basis of the form $\ket{j_{\!f},i_n}$ where $(j_{\!f},i_n)$ are the colors of a spin network $\Delta_2^*$ (dual to the triangulation of physical space) formed by two nodes $n=1,2$  joined by four links  $f=1,...,4$.  
\begin{center}
 \hspace{2em}
\begin{picture}(20,40)
\put(-36,17) {$\Delta_2^*\  =$}
\put(20,20) {\circle{30}}
\put(04,20) {\circle*{4}} 
\put(36,20) {\circle*{4}}  
\qbezier(4,20)(20,33)(36,20)
\qbezier(4,20)(20,7)(36,20)
\end{picture} \hspace{2.3em}\raisebox{18pt}{.}
\end{center} 
The Hamiltonian constraint operator is 
\be
\tilde C\, \psi(j_{\!f}, \iota_t)\  
=\sum_{\epsilon_j=0,\pm 1}
{C}^{\epsilon_{\!f} \iota_t'}_{j_{\!f} \iota_t}\  \psi\!\left(j_{\!f}+\frac{\epsilon_{\!f}}{2}, \iota'_t\right), 
\ee
where the coefficients $C^{\epsilon_j \iota_t'}_{j_{\!f} \iota_t}$ are defined in \cite{Rovelli&Vidotto1}; they  vanish unless $\epsilon_{\!f}\!\!=\!\!0$ for two and only two of the four $j$'s.   This Hamiltonian is not convenient for the spinfoam expansion because it lacks a diagonal term.  However, it is easy to replace it with an operator that has a non-vanishing diagonal term: it suffice to write the holonomy that defines the regularization of the curvature in the spin-one representation instead than the spin-half representation \cite{gaul}. This gives
\be
\tilde C\, \psi(j_{\!f}, \iota_t)\  
=\sum_{\epsilon_j=0,\pm 1}
{C'}^{\epsilon_{\!f} \iota_t'}_{j_{\!f} \iota_t}\  \psi\!\left(j_{\!f}+{\epsilon_{\!f}}, \iota'_t\right), 
\ee
where now the coefficients ${C'}^{\epsilon_j \iota_t'}_{j_{\!f} \iota_t}$ vanish unless $\epsilon_{\!f}\!\!=\!\!0$ for at least two of the four $j$'s. In particular, they are nonvanishing for all $\epsilon_{\!f}$'s equal to zero.    In this form, we can immediately apply the results of the previous section. The regularized transition amplitudes (or the physical scalar product) between two $\ket{j_{\!f},i_n}$ states will be given by 
\ba\label{proggycbis}
\bk{j'_{\!f},i'_n}{P_\delta}{j_{\!f},i_n}&\!\!\!=&\!\!\!\sum_{M=0}^\infty
\sum_{j^{1}_{\!f},i^{1}_n ... j^{M-1}_{\!f},i^{M-1}_n } C_{j'_{\!f},i'_n j^{M-1}_{\!f},i^{M-1}_n}
\nn
&&\!\!\! \!\!\! \!\!\! ... C_{j^{1}_{\!f},i^{1}_n j_{\!f},i_n }
  \ \mu_\delta(j^{1}_{\!f},i^{1}_n ... j^{M-1}_{\!f},i^{M-1}_n),
\ea
where 
\ba\label{eccola22222}
\mu_\delta(j^{1}_{\!f},i^{1}_n ... j^{M-1}_{\!f},i^{M-1}_n)&&=
i^{\scriptscriptstyle M}\,
\tfrac{\delta}{2}\int^{1/\delta}_{-1/\delta}
\dd\tm \
\int_0^t \dd\tm_M  ...  \int_0^{t_2} \dd\tm_1 \nn
&&(e^{i C_{j'_{\!f},i'_n}})^{t-t_{M}}
...
(e^{i C_{j_{\!f},i_n}})^{t_1}\ .
\ea
This expression give the matrix elements of the projection operator in terms of a sum over 
histories of finite length. These histories are sequences of spin networks with the same graph $\Delta_2^*$. The spins can change at most by one at each step.  The intertwiners can change arbitrarily, compatibly with the spins, at each step. The ``vertex" operator is given by the non-diagonal elements of the Hamiltonian constraint
\be
C_{j_{\!f}+{\epsilon_{\!f}},i'_n, j_{\!f},i_n}=\bk{j_{\!f}+{\epsilon_{\!f}},i'_n}{\tilde C}{j_{\!f},i_n}={C'}^{\epsilon_{\!f} \iota_t'}_{j_{\!f} \iota_t}\, .
\ee
We expect that this sum converges rapidly for finite $\delta$.

\vskip3em
\section{Conclusion}

The results that we have obtained are the following.  First, we have introduced a simple way to regularize the spinfoam expression for the projector operator that defines the quantum dynamics in a background independent system. 

Second, we have applied this result for writing a spinfoam version of the ``dipole cosmology".   The main advantage of this procedure is that a scalar field to keep track of the temporal evolution is not needed.   The system has enough degrees of freedom for expressing the dynamics in a fully relational way, without the need of deparameterizing it. 
All physical amplitudes can be expressed as ``transition amplitudes" (or matrix elements of the physical scalar product) among the unconstrained spin network states. For a general discussion of the physical interpretation of these transition amplitudes, see \cite{libro}.

The same technique can in principle be applied to other cosmological models with enough degrees of freedom.  In particular, the application of the same technique for writing a spinfoam formulation of a Bianchi I cosmology will appear elsewhere \cite{CFE}.

\vskip.5cm
\centerline{---}
\vskip.5cm

We thank Ed Wilson-Ewing for many illuminating
discussions, and Matteo Smerlak for a clarifying discussion
on asymptotic expansions. FV acknowledges the
support of the ÓFondazione della RicciaÓ.

\end{document}